\documentclass[aps,prb,twocolumn,showpacs,preprintnumbers,amsmath,amssymb,superscriptaddress]{revtex4}

\usepackage{graphicx, epsfig}
\usepackage{bm}
\usepackage{color}
\usepackage{hyperref}

\usepackage{ulem}

\begin{document}


\title{Infrared Conductivity of Elemental Bismuth under Pressure: Evidence for an Avoided Lifshitz-Type Semimetal-Semiconductor Transition}

\author{N.~P.~Armitage}
\affiliation{Department of
Physics and Astronomy, The Johns Hopkins University, Baltimore, Maryland
21218, USA.}
\affiliation{D\'{e}partement de Physique de la Mati\`{e}re
Condens\'{e}e, Universit\'{e} de Gen\`{e}ve, quai Ernest-Ansermet
24, CH1211 Gen\`{e}ve 4, Switzerland.}
\author{Riccardo Tediosi}
\affiliation{D\'{e}partement de Physique de la Mati\`{e}re
Condens\'{e}e, Universit\'{e} de Gen\`{e}ve, quai Ernest-Ansermet
24, CH1211 Gen\`{e}ve 4, Switzerland.}
\author{F. L\'evy}
\affiliation{D\'{e}partement de Physique de la Mati\`{e}re
Condens\'{e}e, Universit\'{e} de Gen\`{e}ve, quai Ernest-Ansermet
24, CH1211 Gen\`{e}ve 4, Switzerland.}
\author{E. Giannini}
\affiliation{D\'{e}partement de Physique de la Mati\`{e}re
Condens\'{e}e, Universit\'{e} de Gen\`{e}ve, quai Ernest-Ansermet
24, CH1211 Gen\`{e}ve 4, Switzerland.}
\author{L. Forro}
\affiliation{Ecole Polytechnique Federale de Lausanne, Switzerland.}
\author{D. van der Marel}
\affiliation{D\'{e}partement de Physique de la Mati\`{e}re
Condens\'{e}e, Universit\'{e} de Gen\`{e}ve, quai Ernest-Ansermet
24, CH1211 Gen\`{e}ve 4, Switzerland.}

\date{\today}

\begin{abstract}

The application of pressure to elemental bismuth reduces its conduction-valence band overlap, and  results in a semimetal-semiconductor (SMSC) transition around 25 kbar.  This transition is nominally of the topological ``Lifshitz" Fermi surface variety, but there are open questions about the role of interactions at low charge densities.  Using a novel pressure cell with optical access, we have performed an extensive study of bismuth's infrared conductivity under pressure.   In contrast to the expected pure band behavior we find signatures of enhanced interaction effects, including strongly coupled charge-plasmon (plasmaron) features and a plasma frequency that remains finite up to the transition.  These effect are inconsistent with a pure `Lifshitz' bandlike transition.  We postulate that interactions plays a central role in driving the SMSC transition.

\end{abstract}

\pacs{71.45.-d, 78.40.Kc, 78.20.-e, 78.30.-j}
\maketitle

Since the seminal work of Wigner in 1930s, it has been believed that the large relative electron interactions in low density charge systems manifest themselves in a variety of novel phases like electronic crystals, correlated heavy electron fluids, and inhomogeneous states \cite{Wigner34a,Ceperley80a,Spivak09a}.  In principle elemental bismuth is a model system to investigate such physics.  It is a semimetal with small valence and conduction band overlap, three small $L$ point electron Fermi pockets, and a $T$ point hole Fermi pocket, which gives an equal (small) number of both charge species at $E_F$.  A material of much long term interest with many interesting properties \cite{edelman76}, bismuth has recently been found to host a variety of exotic electronic phenomena, including phase transitions at high field to a ``valley-ferromagnetic" state \cite{Lu08a}, signatures of charge fractionalization in the ultraquantum high field limit \cite{Behnia07a}, and strongly coupled electron-plasmon ``plasmaron"  features in the optical spectra \cite{Tediosi07a}.  Recently, it also has been claimed that Bi$_{1-x}$Sb$_x$ alloys are `topological insulators', with robust topologically protected surface states \cite{Hsieh08a}.

\begin{figure}[htbp]
\begin{center}
\includegraphics[width=7.6cm,angle=0]{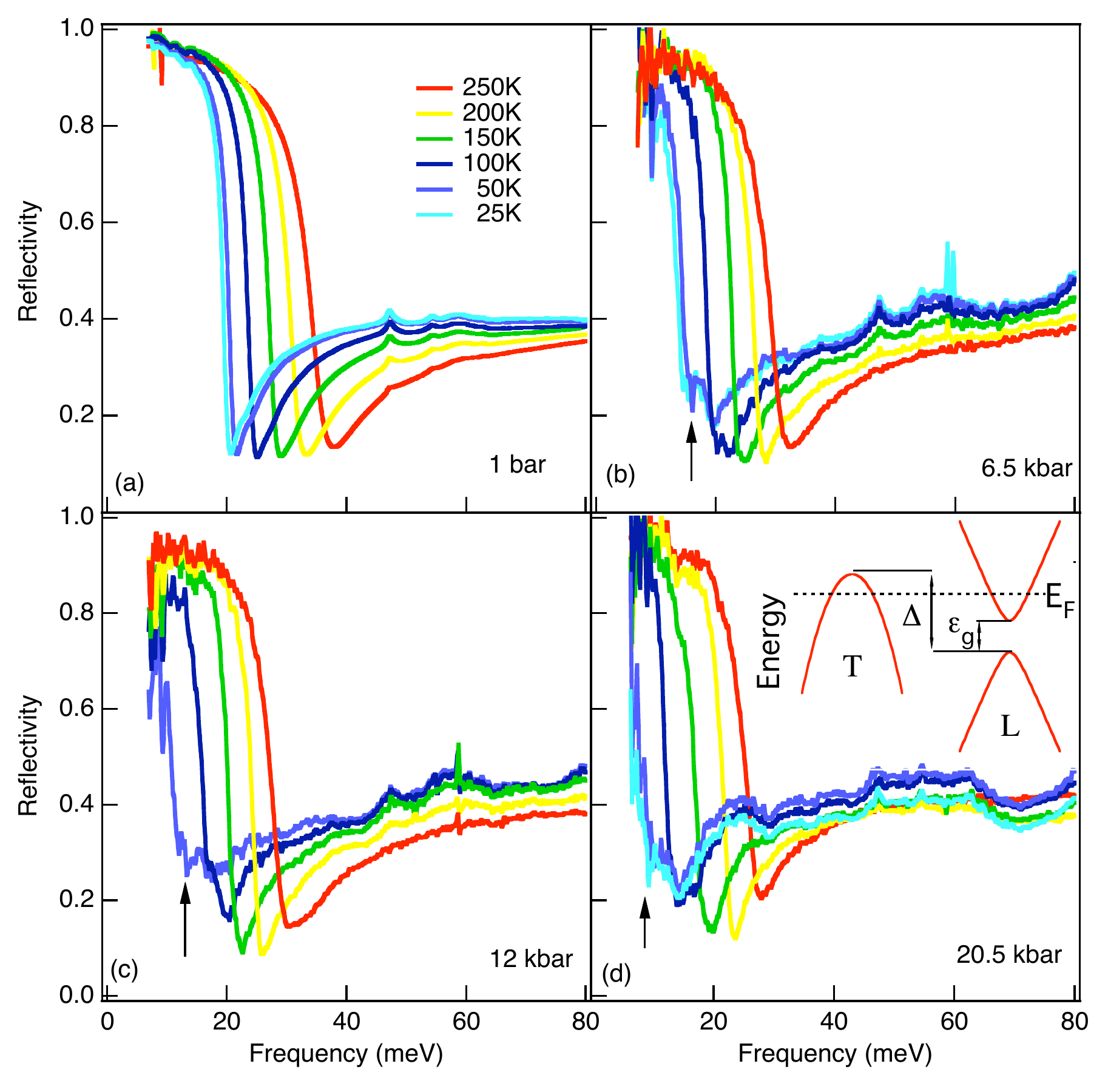}
\caption{(Color) Reflectivity of an effective bismuth-air interface at a number of different pressures.  The small 47 meV feature is an artifact of the IR beam splitter.  Small black arrows denote the ``plasmaron" feature.}
\label{Reflectl}
\end{center}
\end{figure}

Among its remarkable phenomena, bismuth undergoes an interesting semimetal-semiconductor (SMSC) transition with the application of modest pressures ($\approx$ 25 kbar) \cite{balla65}.  This transition is relatively under investigated \cite{Itskevich68a,Kraak82a,balla65}, but unlike more conventionally discussed Mott- or Anderson-style metal-insulator transitions driven by local interactions or disorder respectively, it is believed to be driven by a reduction of the semimetal conduction and valence band overlap $\Delta$ (see Fig. \ref{Reflectl}).  As the overlap decreases, Fermi surfaces shrink and eventually vanish.  Within a non-interacting purely band point of view this is an electronic topological transition of the ``Lifshitz" variety \cite{Lifshitz60a}.  It is unclear however what role is played by electronic correlations on the approach to the transition when charge densities are low.   Do interactions dominate resulting in a strongly correlated liquid state? Perhaps the continuous topological `Lifshitz' band transition is superseded by a first-order discontinuous transition \cite{Kaganov84a} or a phase like an electronic crystal \cite{Wigner34a,Ceperley80a}, excitonic insulator \cite{Halperin68a}, or inhomogeneous state \cite{Spivak09a}.

Bismuth has an $A7$ rhombohedral lattice.  Its exceptional electronic properties originate in the small deviation of this structure from higher symmetry simple cubic.  It can be formed from the successive distortions of two interpenetrating fcc sublattices: first a small relative displacement along the body-diagonal, and then a small rhombohedral shear along the same direction.  The body-diagonal displacement doubles the real space lattice constant and gaps states at the chemical potential and reduces electronic energy in the usual fashion.  In a hypothetical structure with only the Peierls-like body diagonal displacement, the gapping would be complete and the material a semiconductor with zero near-E$_F$ density of states \cite{Shick99a}.   The additional symmetry lowering from finite rhombohedral shear allows valence and conduction band overlap \cite{Shick99a}.  Such considerations allow us to understand the effect of pressure.     Bismuth has a strongly anisotropic compressibility, with the c axis 3 times more elastic than the perpendicular direction\cite{White72a}.   Since the angle associated with the rhombohedral shear determines the unit cell volume, it couples strongly to pressure.   Increased pressure drives the angle back towards the more symmetric cubic value (60$^\circ$) and decreases the band overlap, driving the material through a SMSC transition.

In this Letter, we present the results of an infrared conductivity study of bismuth up to its pressure tuned SMSC transition.  Near the critical pressure we find colossal changes in its optical conductivity, evidence for correlation effects and a plasma frequency that stays finite through the transition showing that this transition is likely not of a pure Lifshitz variety.  Our work also represents an important proof of principle for the technique in which we have demonstrated the ability to perform reliable far-infrared measurements under pressure and at low temperatures without a synchrotron.

The single-crystal sample used in this work was grown by a modified Bridgman-Stockbarger technique\cite{Tediosi07a}.  We confirmed with x-ray diffraction that the mirrorlike cleavage surfaces  were [110] planes perpendicular to the trigonal axis.  Measurements were performed with a standard Bruker 66 IR spectrometer.  Ambient pressure spectra were measured in our usual optical cryostat\cite{Tediosi07a}. Using a novel kerosene filled piston-cylinder cell\cite{Kezsmarki05a} equipped with a large 2 mm diamond window we performed infrared reflection experiments on a sample mounted in contact with the window.  A 2$^\circ$ angular miscut of the diamond allowed us to separate the contribution from the front and back interfaces. The unusually large optical access and a special kinematic mounting for the cell allowed us to measure into far-infrared.  The pressure dependent resistance of an InSb chip was used to determine the pressure at temperatures down to 10 K.

The determination of the complex conductivity $\sigma(\omega)$ or dielectric function $\varepsilon(\omega)$ consists of three main steps \cite{Tediosi08a}. First, using a rigorous calibration procedure, the signal was corrected for the stray light reflected at the diamond-vacuum interface. Next, the data are combined with the reflectivity measured outside the pressure cell to provide the absolute reflectance of the diamond-sample interface at room temperature and for each pressure. Finally, the relative change of the intensity as a function of temperature gives the reflectivity $R(\omega,T)$.   The complex optical constants are generated via a Kramers-Kronig consistent fitting procedure of $R(\omega)$  constrained by ambient pressure ellipsometry between 0.7 end 4.5 eV\cite{alexey}.

\begin{figure}[htbp]
\begin{center}
\vspace{3mm}
\includegraphics[width=8.0cm,angle=0]{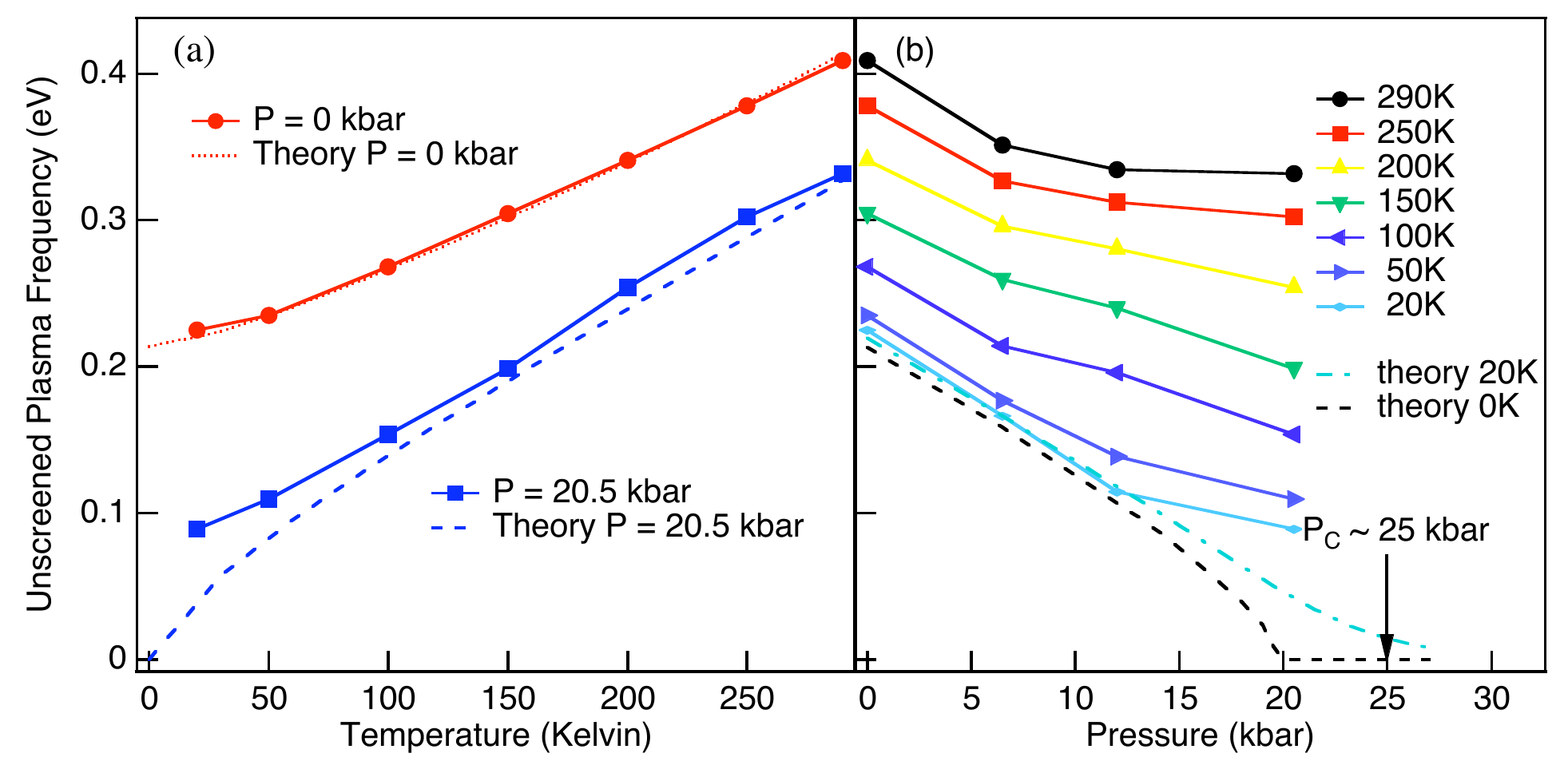}
\caption{(Color) (a)  Temperature dependence of the plasma frequency at ambient pressure and at 20.5 kbar.  (b)  Plasma frequency as a function of pressure at various temperatures.}
\label{PlasmaFreq}
\end{center}
\end{figure}

In Fig. \ref{Reflectl} we show the effective reflectivity of the bismuth-air interface at different pressures and temperatures.   As discussed previously \cite{Tediosi07a}, the reflectivity exhibits a sharp plasma edge that decreases rapidly with temperature.   In the optical response of simple metals, the plasma edge is exhibited when $\varepsilon_1(\omega)=0$, which is satisfied self-consistently at the screened plasma frequency  $\tilde{\omega}_p = \omega_p / \sqrt{\varepsilon_\infty(\tilde{\omega}_p )}$.   Here $\varepsilon_\infty$ is the contribution from everything except the Drude term and $\omega_p^2$ is the Drude spectral weight $4 \pi n e^2/m_b$ ($n$ is the charge density and $m_b$ is the band mass).  $\tilde{\omega}_p$ occurs at anomalously small values, due to both its low charge density and exceptionally large  $\varepsilon_\infty \approx 85-95$.  The decreasing plasma frequency as a function of temperature reflects the freezing out of charges at low temperature (Fig. \ref{PlasmaFreq}(a)).

Superficially the effect of pressure looks similar to the effect of temperature.  However in this case the plasma edge decreases due to a smaller band overlap.  The behavior is summarized in Fig. \ref{PlasmaFreq}(b), where we plot as a function of pressure at different temperatures the plasma frequency $\omega_p$  of the Drude component from a Drude-Lorentz fit to the reflectivity.   At low pressure the data is roughly consistent with a free charge density extrapolating to zero near the known critical pressure.   However at higher pressures, the experimental data extrapolate to a finite value at the transition point.   We will return to this point (and the overlaid theoretical curves) below.

\begin{figure}[htbp]
\begin{center}
\includegraphics[width=8cm,angle=0]{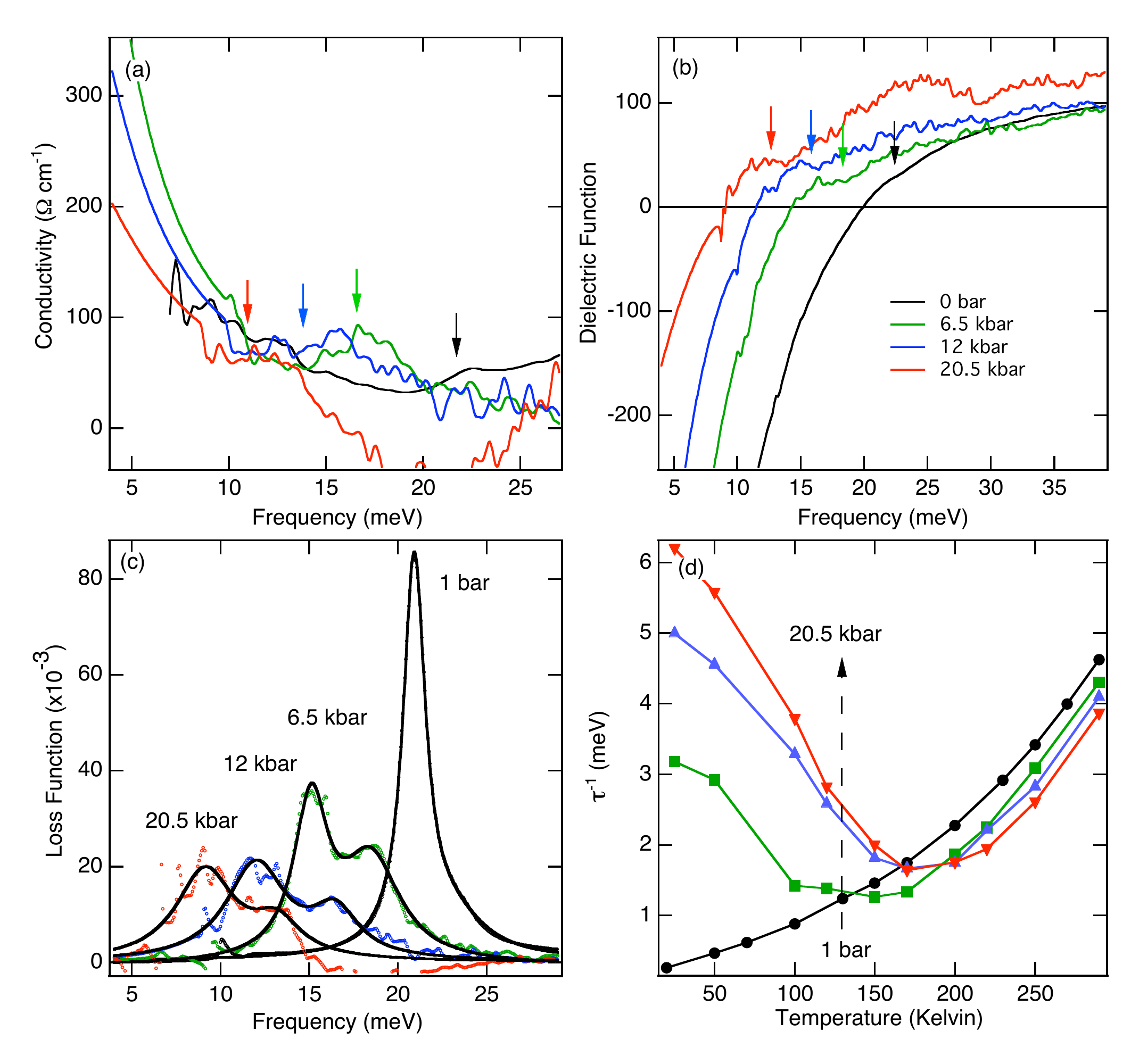}
\caption{(Color) (a)  Real optical conductivity for 6.5, 12, and 20.5 kbar at 20K.  Arrows denote position of `plasmaron' absorption.  The small negative shift in the 20.5 kbar spectra is an artifact of the calibration procedure.  (b)  Real part of the dielectric function.  (c)  Low temperature (20K) loss function. (d) Temperature dependence of the scattering rate as a function of temperature at different pressures.   Data below 7 meV is an extension of the fit in the Kramers-Kronig analysis.}
\label{LF}
\end{center}
\end{figure}

In Figs. \ref{LF}(a) and (b) we plot the real parts of the low temperature (20 K) optical conductivity and dielectric function at different pressures.  Although the gross aspects of the spectra are as expected for a metallic system with a small number of free charges, a more detailed analysis shows a number of interesting fine features.   At 6.5 kbar we observe a plateau in the conductivity and an inflection point in the dielectric function around 18 meV.  These features systematically move to lower energies as the pressures increases (see arrows).  The inflection point in the real part of the dielectric function also becomes progressively more pronounced as the pressure increases. This behavior may anticipate a second zero-crossing of $\varepsilon_1$ which could occur in proximity to the SMSC transition.

These features, which are found in the vicinity of the screened plasmon frequency, are evidence for the  coupled electron-plasmon ``plasmaron" feature that we have previously observed in the ambient pressure spectra \cite{Tediosi07a}.   This is an electron-boson interaction similar in character to electron-phonon or electron-magnon interactions, except here the collective mode is purely electronic.  As noted by Lundqvist \cite{lundqvist1} the plasmon pole contribution to the quasiparticle self-energy has a similar form to that of the polaron problem.   Indeed the term plasmaron follows from this analogy.  Note that such effects should only be visible in the conductivity in a system which breaks Galilean invariance, which here is afforded by the simultaneous presence of electrons and holes.  Such coupling effects are particularly significant here due to the anomalously low plasmon frequency.  They become even more pronounced as the charge density vanishes close to the SMSC, showing the increased role of interactions at low charge densities.  Note that these plasmaron features are visible in the raw reflectivity curves themselves;  a structure additional to the usual plasma minimum develops near the plasma edge at high pressures and low temperature as denoted by the arrows in Fig. \ref{Reflectl}.

As may be expected, the plasmaron features show up dramatically in the loss function Im$ \frac{-1}{\varepsilon(\omega)}$ itself, as an additional peak or shoulder near the usual plasmon peak as shown in Fig.~\ref{LF}(c).  With increasing pressure, spectral weight shifts from the main plasmon peak to the ``plasmaron'' derived peak.  Such shifts of spectral weight from primary to satellite peaks are consistent with increasing interactions exhibiting themselves in an enhancement of the optical mass as one approaches the SMSC transition.

This plasmaron coupling manifests itself directly in the low energy physics.   In Fig. \ref{LF}(d) we plot the Drude scattering rate determined from the Drude-Lorentz fits, which at finite pressure has a distinct nonmonotonic form.   It is interesting to note that it has the same qualitative dependence as the dc resistivity itself \cite{balla65}.   From the dc data alone, it is not clear if a nonmonotonic resistivity could be caused by the competing effects of a decreasing scattering rate and decreasing carrier density.   Here we show that unusual nonmonotonic dependencies exist in the scattering rate itself.  This shows that interactions in the form of plasmon coupling plays an increasingly important role as one approaches the nominal Lifshitz transition.  Also note that the high temperature scattering rate $decreases$ at high pressures.

As noted above and shown in Fig. \ref{PlasmaFreq}, although the low temperature plasmon frequency is significantly suppressed as one approaches the nominal critical pressure, a reasonable extrapolation  does not appear to go to zero at  $p_c$ as would be expected if the free carrier density vanishes $\grave{a}$ $la$ Lifshitz.  In Fig. \ref{PlasmaFreq}, overtop the experimental data we plot a band calculation of the pressure and temperature dependent plasma frequency from the Drude spectral weight.   The calculation was performed taking into account the three relevant charge species: (1) heavy holes ($hh$) at the $T$ point with a standard massive dispersion ($\epsilon_k^{hh}=\hbar^2k^2/m_k^{hh}$), (2)  light electrons ($le$) and (3) their light hole ($lh$) counterparts at the threefold degenerate $L$ point, for which the energy dispersion is: $\epsilon_k=\Delta-\epsilon_g/2\pm\sqrt{\epsilon_g^2/4+\hbar^2|\vec{k} \cdot \vec{v_F}|^2}$ where $\epsilon_g$ is the direct gap between the two bands.  The effective mass-tensor of the light carriers near the gap is defined by $\vec{m}=\epsilon_g/(2\vec{v}_F^2)$.  Band parameters are given in Table~ \ref{table_parameters}.  The SMSC transition is governed by the band overlap  $(-\Delta)$ of the heavy hole and light electron bands.  By adjusting the chemical potential the calculation fulfills charge neutrality: $n_{hh}+n_{lh}=n_{le}$. The plasma frequency $\omega_p$ follows from the general relation: $\omega_p^2=4\pi e^2\sum_j\left[\sum_kn^j_k/m_k^j\right]$ where the $j$ is a band index and $\hbar^2/m_k=-\partial^2\epsilon_k/\partial k^2$. 
 
 \begin{table}[htbp]
\begin{center}
\begin{tabular}{|l|c|}
\hline Heavy hole masses & $m_a=0.064$, $m_b=0.064$, $m_c=0.69$\\
\hline
Light particle velocities & $v_F^1=1080 $, $v_F^2=71.1$, $v_F^3=669 $\\
\hline
Band overlap $\Delta$&$\Delta=\Delta^0\times(1-0.07T/\Delta^0-p/p_c)$ \\
&$\Delta^0=-40$~meV, $p_c=20$~kbar\\
\hline
Direct gap $\epsilon_g$&$\epsilon_g=\epsilon_g^0(1+2.5e^{-4}T^2/\epsilon_g^0)$ \\
&$\epsilon_g^0=15$~meV\\
\hline
\end{tabular}
\caption{Parameters used for the plasma frequency calculation. ``$0$" stands for zero temperature and pressure; masses are in units of the free electron mass; $a$,$b$,$c$ are respectively the binary, bisectrix and trigonal axis;  the (2,3) plane is  tilted $6.23^o$ relative to (b,c); velocities are in $km/s$.  Parameters were adopted from Hall, magneto-optic, cyclotron resonance and Shubnikov-de Haas oscillation measurements \cite{Norin77a,Mendez81a,Smith64a}.
}
\label{table_parameters}
\end{center}
\end{table}

At zero pressure, the calculation fits $\omega_p$ exceedingly well over the entire temperature range while at 20.5 kbar there is a substantial disagreement at low temperature (Fig. \ref{PlasmaFreq}(a)).  From Fig. \ref{PlasmaFreq}(b) the discrepancy is enhanced as the SMSC transition is approached. The calculation predicts the SMSC transition to occur at 20 kbar, which is completely missed by the experimental data.  Note that there is a reasonable amount of uncertainty in the pressure dependence of the band parameters \cite{Norin77a,Mendez81a,Smith64a} and so the calculation should not be taken too literally, but in any case the experimental $\omega_p$ does not appear to extrapolate to zero at the known $p_c$ near 25 kbar.

The appearance of correlation spectral features and finite $\omega_p$  on the approach to the SMSC transition shows that pure Lifshitz-like behavior is superseded by some other - presumably interaction driven - effects.  A naive estimate of the effective dimensionless interaction parameter $r_s$ at ambient pressure is small if one considers the large $\varepsilon_\infty$ at the screened plasma frequency and a small effective mass (for the holes $r_s  \approx 0.2$).   However, the appropriate $\varepsilon_\infty$ for the bare interaction strength is likely not the $q=0$ one relevant at the plasma edge, but instead one at short distances that is likely much smaller.  This is reasonable when one considers that the typically cited large $\varepsilon_\infty$ derives from low energy, but highly dispersive interband transitions at the  $L$ point \cite{Alstrom81a} and moreover, that the dominant interaction in these systems has a short-range excitonic character.  A larger $r_s$ means that near the transition interactions should become significant.

A number of possibilities exist for the physics that supersedes the Lifshitz transition:
 
(i) At relatively high density an electron-hole plasma can break up into charge rich and charge poor regions (but still locally neutral with $n=p$) in a manner similar to electron-hole droplets \cite{Jeffries88a} or ``microemulsions"\cite{Spivak09a} in semiconductors. The charge poor regions may be insulating as described below, while the charge rich regions remain metallic. In this scenario, the transition at $p_c$ would essentially be one of percolative character.    
 
(ii) At low density the holes can form a crystal while the electrons remain liquid, provided that their mass ratio exceeds approximately 100 \cite{Bonitz05a}. While in bismuth this condition is met along the c-axis, where the mass ratio is about 200, the ratio is smaller in other directions. Whether or not a hole crystal can form requires better understanding of the role of the strong mass-anisotropy. 
 
(iii) When the material is driven close to the pressure where $\Delta$ becomes positive, electron-hole bound states may form rendering the material insulating. Excitonic correlations\cite{Halperin68a,Yamaji06a,Bronold07a} are believed to play a role at similar densities in TmSe$_{0.45}$Te$_{0.55}$ \cite{Wachter04a}.
 
Although throughout our experimental range the system clearly remains metallic, plasmaron features could be the manifestation of enhanced  correlations on the approach to the transition \cite{Sayakanit85}. Despite these clear signatures of interaction effects  there is clearly more work to be done on this interesting system. We hope that our work is stimulating in that regard.

We would like to thank  K. Behnia, A. MacDonald, O. Vafek, and Z. Wang for helpful discussions and R. Gaal for technical support.  This work is supported by the SNSF through Grant No. 200020-125248 and the National Center of Competence in Research (NCCR) MaNEP.  NPA was supported by NSF DMR-0847652 and the NSF International Research Fellows Program.

\bibliography{BismuthBib_etal}%

\end{document}